\documentclass[a4paper,english,runningheads]{llncs}
\usepackage[T1]{fontenc}
\usepackage[latin9]{inputenc}
\usepackage{color}
\usepackage{units}
\usepackage{amsmath}
\usepackage{amssymb}
\usepackage[all]{xy}

\makeatletter

\usepackage[all]{xy}

\usepackage{babel}

\makeatother

\usepackage{babel}
\begin{document}

\title{A Qualified Kolmogorovian Account of Probabilistic Contextuality}

\titlerunning{Probabilistic Contextuality}

\author{Ehtibar N. Dzhafarov\textsuperscript{1}%
\thanks{Corresponding author.%
}\and Janne V. Kujala\textsuperscript{2}}

\authorrunning{E. N. Dzhafarov and J. V. Kujala}

\institute{\textsuperscript{1}Purdue University\\
 ehtibar@purdue.edu\\
 \textsuperscript{2}University of Jyväskylä\\
 jvk@iki.fi\\
 }

\toctitle{A Qualified Kolmogorovian Account of Probabilistic Contextuality}

\tocauthor{E. N. Dzhafarov and J. V. Kujala}
\maketitle
\begin{abstract}
We describe a mathematical language for determining all possible patterns
of contextuality in the dependence of stochastic outputs of a system
on its deterministic inputs. The central principle (contextuality-by-default)
is that the outputs indexed by mutually incompatible values of inputs
are stochastically unrelated; but they can be coupled (imposed a joint
distribution on) in a variety of ways. A system is characterized by
a pattern of which outputs can be ``directly influenced'' by which
inputs (a primitive relation, hypothetical or normative), and by certain
constraints imposed on the outputs (such as Bell-type inequalities
or their quantum analogues). The set of couplings compatible with
these constraints determines the form of contextuality in the dependence
of outputs on inputs.

\end{abstract}

\section{Introduction}

In this paper we describe a language for analyzing dependence of \emph{stochastic
outputs} of a system on \emph{deterministic inputs}. This language
applies to systems of all imaginable kinds: quantum physical, macroscopic
physical, biological, psychological, and even purely mathematical,
created on paper. The notion of ``dependence,'' as well as related
to it notions of ``influence,'' ``causality,'' and ``context''
may have different meanings in different areas. Even if not, we do
not know how to define them. We circumvent the necessity of designing
these definitions by simply accepting that some inputs are connected
to some outputs by arrows called \emph{direct} \emph{influences}.
We ignore the question of how these direct influences are determined,
except for a certain necessary condition they must satisfy (\emph{marginal
selectivity}). A system is also characterized by certain \emph{constraints}
imposed on the joint distribution of its outputs across different
inputs. A prominent example when both direct influences and constraints
are justified by a well-developed theory is the EPR paradigm in quantum
physics, where it is assumed that measurement settings for a given
particle directly affect measurement outcomes in that particle only,
and the joint distributions of the measurement outcomes on different
particles satisfy certain inequalities or parametric equalities. If
these constraints can be accounted for entirely in terms of the posited
direct influences, the system can be viewed as ``contextless.''
If this is not the case, we characterize \emph{probabilistic contexts}
by studying the deviations from the contextless behavior exhibited
by the system. 

Whether one deals with quantum contextuality or thinks of contextuality
beyond even quantum bounds, our approach does squarely remains within
the domain of the classical probability theory, which we refer to
as \emph{Kolmogorovian}. A caveat for using this attribution is that
we do not mean the ``naive'' Kolmogorovian theory in which all random
variables are thought of as defined on a single sample space (equivalently,
as functions of a single random variable). Such a notion is no more
tenable than the ``set of all sets'' of the naive set theory. The
qualified Kolmogorovian approach we adopt is based on the principle
of \emph{contextuality-by-default: }
\begin{quotation}
\emph{any two random variables recorded under mutually exclusive conditions
are stochastically unrelated}, \emph{defined on different sample spaces}. 
\end{quotation}
This is a radical version of views previously expressed in the literature,
e.g., in Khrennikov, 2008a-b, where it is traced back to Andrei Kolmogorov
himself and even to George Boole. Our emphasis, however, is on the
fact that any set of stochastically unrelated variables (but never
``all of them'') can be \emph{coupled}, or imposed a joint distribution
upon, in many different ways (Thorisson, 2000). In particular, the
identity coupling is sometimes (but not always) possible, in which
the two random variables defined under mutually exclusive conditions
and ``automatically'' (by default) labeled as different and stochastically
unrelated, merge into one and the same random variable. 

The basics of this approach are presented in Section 2. In Sections
3 and 4 we use it to investigate contextual influences with respect
to a given pattern of direct influences. The theory and notation there
closely follows Dzhafarov and Kujala (2013a). The departure point
is that since different treatments (combinations of input values)
are mutually exclusive, the joint distributions of the outputs corresponding
to them, according to the principle of contextuality-by-default, are
stochastically unrelated. We then consider all possible ways of coupling
them across different treatments. From each such a coupling we extract
stochastic relations that are ``hidden,'' principally unobservable,
because they correspond to outputs obtained under different treatments.
We focus on the special kind of these hidden relations, those between
random variables that share the same pattern of direct influences.
We call these hidden relations \emph{connections}. Given a certain
constraint imposed on the system by a theory or empirical observations,
we pose the question of what connections imply (or force) this constraint
and what connections are implied by (or compatible with) it. Taken
over all possible couplings, these relations between connections and
constraints characterize the type of contextuality exhibited by the
system. This view of contextuality is different from the existing
approaches (Khrennikov, 2009; Laudisa, 1997).

\section{Probability Theory: Multiple Sample Spaces}

Given two probability spaces, $\left(S,\Sigma,p\right)$ and $\left(S_{A},\Sigma_{A},p_{A}\right)$,
with standard meaning of the terms, a \emph{random variable} is defined
as a $\left(\Sigma,\Sigma_{A}\right)$-measurable function $A:S\rightarrow S_{A}$
subject to
\begin{equation}
p_{A}\left(X\right)=p\left(A^{-1}\left(X\right)\right),\label{eq:measurable}
\end{equation}
for any $X\in\Sigma_{A}$. The probability space $\left(S,\Sigma,p\right)$
is usually called a \emph{sample space}, and we will refer to $\left(S_{A},\Sigma_{A},p_{A}\right)$
as the \emph{distribution} of $A$. The sample space itself is a distribution
of the random variable $R$ (let us call it a \emph{basic variable})
which is the $\left(\Sigma,\Sigma\right)$-measurable identity function,
$x\mapsto x$, $x\in\Sigma$. Any random variable $A$ defined on
this sample space can also be presented as a function $A=f\left(R\right)$,
and (\ref{eq:measurable}) can be written as
\begin{equation}
p_{A}\left(X\right)=\Pr\left[A\in X\right]=\Pr\left[R\in f^{-1}\left(X\right)\right],\label{eq:measurable'}
\end{equation}
for any $X\in\Sigma_{A}$. 

Let $\left(A^{k}=f_{k}\left(R\right):k\in K\right)$ be a sequence%
\footnote{The term \emph{sequence} in this paper is used in the generalized
meaning, as any indexed family, a function from an index set into
a set. Index sets need not be countable.%
} of random variables, all functions of one and the same basic variable
$R$, with $A^{k}$ distributed as $\left(S^{k},\Sigma^{k},p_{k}\right)$.
Then $A=\left(A^{k}:k\in K\right)=f\left(R\right)$ too is a random
variable that is a function of $R$, with the distribution
\begin{equation}
\left(S_{A}=\prod_{k\in K}S^{k},\Sigma_{A}=\bigotimes_{k\in K}\Sigma^{k},p_{A}\right).\label{eq:joint}
\end{equation}
Here, $\bigotimes_{k\in K}\Sigma^{k}$ is the minimal sigma-algebra
containing sets of the form $X^{k}\times\prod_{i\in K-\left\{ k\right\} }S^{i}$
for all $X^{k}\in\Sigma^{k}$, and $p_{A}$ is defined by (\ref{eq:measurable'}),
with
\begin{equation}
f^{-1}\left(X\right)=\left\{ x\in S:\left(f_{k}\left(x\right):k\in K\right)\in X\right\} .
\end{equation}

The distribution of $A$ can also be given by (\ref{eq:joint}) with
no reference to its sample space, or basic variable. It can be viewed
as a \emph{joint distribution} of the components of a sequence $A=\left(A^{k}:k\in K\right)$,
such that, for any nonempty $K'\subset K$, the subsequence $A'=\left(A^{k}:k\in K'\right)$
is a random variable distributed as
\begin{equation}
\left(S_{A'}=\prod_{k\in K'}S^{k},\Sigma_{A'}=\bigotimes_{k\in K'}\Sigma^{k},p_{A'}\right),
\end{equation}
with
\begin{equation}
p_{A'}\left(X\right)=p_{A}\left(X\times\prod_{k\in K-K'}S^{k}\right),
\end{equation}
for any $X\in\Sigma_{A'}$. The distribution $\left(S^{k},\Sigma^{k},p_{k}\right)$
of a single $A^{k}$ is determined by that of the one-element subsequence
$\left(A^{k}\right)$ in the obvious way. All the random variables
$A^{k}$ obtained in this way from $A$ can be viewed as functions
on one and the same basic variable, e.g., $R=A$ itself. 

We see that the relation ``are jointly distributed'' is synonymous
to the relation ``are functions of one and the same basic variable.''
But clearly there cannot be a single basic variable of which all imaginable
random variables are functions. This is obvious from the cardinality
considerations alone, as random variables may have arbitrarily large
sets of possible values. But this is true even if one confines consideration
to all imaginable random variables with any given distribution, provided
it is not concentrated at a point. Let, e.g., $\mathcal{B}$ be a
class (not necessarily a set) of all functions of $R$ that are Bernoulli
(0/1) variables with equiprobable values. That is, each $B\in\mathcal{B}$
is a function $f\left(R\right)$ with $f:S\rightarrow\left\{ 0,1\right\} $,
such that $\Pr\left(R\in f^{-1}\left(\left\{ 0\right\} \right)\right)=\nicefrac{1}{2}$.
Consider a Bernoulli variable $B^{*}$ with equiprobable values such
that for any $B\in\mathcal{B}$,
\begin{equation}
\Pr\left(B=0,B^{*}=0\right)=\nicefrac{1}{4}.
\end{equation}
Then $B^{*}$ cannot be a function of $R$ because it is independent
of (hence is not the same as) any of the elements of $\mathcal{B}$.
If needed, however, one can redefine the basic variable, e.g., as
$R^{*}=\left(R,B^{*}\right)$, with independent $R$ and $B^{*}$,
so that all elements of $\mathcal{B}\cup\left\{ B^{*}\right\} $ become
functions of $R^{*}$.

This simple demonstration shows that the Kolmogorovian approach to
probability is not represented by a single sample space with measurable
functions on it. Rather the true picture is an ``open-ended'' class
(definitely not a set) of basic variables that are \emph{stochastically
unrelated} to each other, each with its own class of random variables
defined as its functions: schematically,

\medskip{}

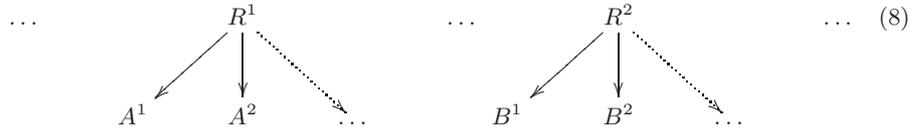
\begin{footnotesize}

\begin{equation}
\xymatrix{\ldots &  & R^{1}\ar[d]\ar[dl]\ar@{.>}[dr] &  & \ldots\\
 & A^{1} & A^{2} & \ldots
}
\xymatrix{ & R^{2}\ar[d]\ar[dl]\ar@{.>}[dr] &  & \ldots\\
B^{1} & B^{2} & \ldots
}
\end{equation}
\end{footnotesize}

\medskip{}

\protect{\noindent}If necessary, using some \emph{coupling scheme}
as discussed below, any sequence of stochastically unrelated basic
variables $\left(R^{k}:k\in K\right)$ can be redefined into a random
variable $H=\left(H^{k}:k\in K\right)$ such that $H^{k}$ and $R^{k}$
are identically distributed for all $k$. This amounts to considering
all individual $R^{k}$, as well as their functions, as functions
of $H$. But this procedure is not unique, and it cannot be performed
for ``all random variables.'' 

The contextuality-by-default principle requires that any two random
variables conditioned upon mutually exclusive values of some third
variable are stochastically unrelated. Indeed, there is never a unique
way for coupling their realizations. A simple example: I flip a coin
and depending on the outcome weigh one of two lumps of clay, lump
1 (if ``heads'') or lump 2 (if ``tails''). The random variables
$A=$``weight reading for lump 1'' and $B=$``weight reading for
lump 2'' do not a priori possess a joint distribution because there
is no privileged way of deciding whether a given value of $A$ \emph{co-occurs}
with a given value of $B$. If necessary, however, such a co-occurrence
(or coupling) scheme can always be constructed. For instance, one
can list the values of $A$ and $B$ chronologically and then couple
the $n$th realization of $A$ with the $n$th realization of $B$
($n=1,2,\ldots$). Or one could rank-order the values of $A$ and
$B$ and couple the realizations of the same quantile rank (this would
create positive correlation between the variables) or of the complementary
ranks (negative correlation). One cannot say that one way of paring
is better justified than another, each one represents ``a point of
view'' and creates its own joint distribution of $A$ and $B$.

\section{All Possible Couplings Approach}

Consider a sequence of random variables $A=\left(A_{\phi}:\phi\in\Phi\right)$.
The elements of $\Phi$ are called \emph{(allowable) treatments}.
Two distinct treatments $\phi,\phi'$ are mutually exclusive, so $A_{\phi}$
and $A_{\phi'}$ are stochastically unrelated. This means that $A$
is not a random variable.

Let there be a sequence of nonempty sets $\alpha=\left(\alpha^{k}:k\in K\right)$
such that $\Phi\subset\prod_{k\in K}\alpha^{k}$. This means that
every treatment is a sequence $\phi=(x^{k}:k\in K)$, with $x^{k}\in\alpha^{k}$.
The sets $\alpha^{k}$ are called \emph{inputs}, and their elements
$x^{k}$ \emph{input values}. Note that generally $\Phi\not=\prod_{k\in K}\alpha^{k}$,
that is, not all possible combinations of input values form treatments
(hence the adjective ``allowable''). 

For every treatment $\phi$, let the random variable $A_{\phi}$ be
a sequence of jointly distributed random variables $A_{\phi}=\left(A_{\phi}^{\ell}:\ell\in L\right)$.
For each $\ell$, the sequence $A^{\ell}=\left(A_{\phi}^{\ell}:\phi\in\Phi\right)$
is called an \emph{output}. Its element $A_{\phi}^{\ell}$ can then
be referred to as \emph{output $A^{\ell}$ at treatment $\phi$} (or
simply output $A_{\phi}^{\ell}$, when this does not create confusion).
Note that $A^{\ell}$ is not a random variable, because its components
are stochastically unrelated. 

We postulate that, for every input $\alpha^{k}$ and every output
$A^{\ell}$, either $\alpha^{k}$ \emph{directly influences} $A^{\ell}$,
and we write $A^{\ell}\leftarrow\alpha^{k}$, or this is not true,
$A^{\ell}\not\leftarrow\alpha^{k}$. This relation is treated as primitive.
Its intuitive meaning can be different in different applications.
The only constraint imposed on this relation, \emph{(complete) marginal
selectivity}, is as follows (Dzhafarov, 2003). Let index subsets $I\subset K$
and $J\subset L$ be such that if $A^{\ell}\leftarrow\alpha^{k}$
for some $\ell\in J$ then $k\in I$. That is, no input belonging
to $\left(\alpha^{k}:k\in K-I\right)$ directly influences any output
belonging to $\left(A^{\ell}:\ell\in J\right)$. Let $\phi=(x^{k}:k\in K)$
and $\phi'=(y^{k}:k\in K)$ be any allowable treatments such that
\begin{equation}
\phi|I=(x^{k}:k\in I)=(y^{k}:k\in I)=\phi'|I.
\end{equation}
The slash here indicates restriction of a function (sequence) on a
subset of arguments (indices). Marginal selectivity means that under
these assumptions 
\begin{equation}
\left(A_{\phi}^{k}:k\in J\right)\sim\left(A_{\phi'}^{k}:k\in J\right),
\end{equation}
where $\sim$ means ``has the same distribution as.'' In other words,
the joint distribution of a subset of outputs does not depend on inputs
that do not directly influence any of these outputs. This does not
mean, however, that these inputs, $\left(\alpha^{k}:k\in K-I\right)$,
can be ignored altogether when dealing with $\left(A^{\ell}:\ell\in J\right)$:
generally, this will not allow one to account for its stochastic relation
to other outputs, $\left(A^{\ell}:\ell\in L-J\right)$.

By appropriately redefining the inputs the relation of ``being directly
influenced by'' can always be made bijective: each output is directly
influenced by one and only one input. The procedure is easier to illustrate
on an example. Let the diagram of direct influences be

\emph{
\begin{equation}
\xymatrix{\alpha^{1}\ar[d]\ar[dr] & \alpha^{2}\ar[d]\ar[dl]\ar[dr] & \alpha^{3}\ar[drr]\ar[dl]\\
A^{1} & A^{2} & A^{3} & A^{4} & A^{5}
}
\end{equation}
}Assume, for simplicity, that all combinations of input values are
allowable, $\Phi=\alpha^{1}\times\alpha^{2}\times\alpha^{3}$. Then
the redefined inputs are as shown:

\emph{
\begin{equation}
\xymatrix{\beta^{1}=\alpha^{1}\times\alpha^{2}\ar[d] & \beta^{2}=\alpha^{1}\times\alpha^{2}\times\alpha^{3}\ar[d] & \beta^{3}=\alpha^{2}\ar[d] & \beta^{4}=\left\{ .\right\} \ar[d] & \beta^{5}=\alpha^{3}\ar[d]\\
A^{1} & A^{2} & A^{3} & A^{4} & A^{5}
}
\end{equation}
}The set $\left\{ .\right\} $ represents a dummy (single-valued)
input, it should be paired with any output that is not directly influenced
by any inputs. The rest of the redefinition should be clear. The set
of allowable treatments is redefined into a new set $\Psi$, which
is not the Cartesian product of the new inputs but rather a proper
subsequence thereof: e.g., if $\beta^{2}$ attains the value $\left(x^{1},x^{2},x^{3}\right)$,
then the only treatment allowable is
\begin{equation}
\left(\left(x^{1},x^{2}\right),\left(x^{1},x^{2},x^{3}\right),x^{2},.,x^{3}\right).
\end{equation}

We assume from now on that the direct influences are defined in a
bijective form: $\alpha=\left(\alpha^{k}:k\in K\right)$, $\Phi\subset\prod_{k\in K}\alpha^{k}$,
$A_{\phi}=\left(A_{\phi}^{k}:k\in K\right)$, $A^{k}\leftarrow\alpha^{k}$
for every $k\in K$, and there are no other direct influences. 

Let us return to the sequence of random variables%
\footnote{In (\ref{eq:A}) and subsequently we are conveniently confusing differently
grouped subsequences, such as $\left(A,B,C\right)$, $\left(\left(A,B\right),C\right)$,
$\left(A,\left(B,C\right)\right)$.%
}
\begin{equation}
A=\left(A_{\phi}:\phi\in\Phi\right)=\left(A_{\phi}^{k}:k\in K,\phi\in\Phi\right),\label{eq:A}
\end{equation}
with stochastically unrelated components. Consider a \emph{complete
coupling} for $A$,
\begin{equation}
H=\left(H_{\phi}^{k}:k\in K,\phi\in\Phi\right),\label{eq:complete coupling}
\end{equation}
a random variable (that is, its components are jointly distributed)
such that
\begin{equation}
H_{\phi}=\left(H_{\phi}^{k}:k\in K\right)\sim\left(A_{\phi}^{k}:k\in K\right)=A_{\phi}.\label{eq:condition for complete coupling}
\end{equation}
Such a random variable $H$ always exists. It suffices, e.g., to consider
every element of $H_{\phi}$ to be stochastically independent of every
element in $H_{\phi'}$, for all $\phi\not=\phi'$. But generally,
the complete couplings $H$ for a given $A$ can be chosen arbitrarily,
except for the defining requirement (\ref{eq:condition for complete coupling}). 

Our approach consists in thinking of $H$, in addition to (\ref{eq:condition for complete coupling}),
in terms of ``connections'' it contains, by which we understand
couplings for sequences of random variables that are indexed by different
treatments sharing the same pattern of direct influences. Consider,
e.g., the components $A_{\phi}^{k}$ for all $\phi$ whose $k$th
element equals a given value $\phi\left(k\right)=x$. This subsequence
can be written as 
\begin{equation}
A_{x}^{k}=\left(A_{\phi}^{k}:\phi\in\Phi,\phi\left(k\right)=x\right).
\end{equation}
Since $A^{k}\leftarrow\alpha^{k}$ only, all random variables $A_{\phi}^{k}$
are directly influenced by the same input value. Let
\begin{equation}
C_{x}^{k}=\left(C_{\phi}^{k}:\phi\in\Phi,\phi\left(k\right)=x\right)\label{eq:example connection}
\end{equation}
be a coupling for $A_{x^{k}}^{k}$. This means that if $\phi\left(k\right)=x$,
\begin{equation}
C_{\phi}^{k}\sim A_{\phi}^{k},
\end{equation}
and it follows from the marginal selectivity property that the distribution
of $C_{\phi}^{k}$ across all $\phi$ with $\phi\left(k\right)=x$
remains unchanged (and equal to the distribution of $A_{\phi}^{k}$).
There can be many joint distributions of (\ref{eq:example connection})
with this property. One possibility is that $C_{x}^{k}$ is an \emph{identity
coupling}, meaning that for any two $C_{\phi}^{k},C_{\phi'}^{k}$
in (\ref{eq:example connection}), 
\begin{equation}
\Pr\left(C_{\phi}^{k}=C_{\phi'}^{k}\right)=1.\label{eq:identity for reduced}
\end{equation}
If this is assumed for all $k\in K$ and $x\in\alpha^{k}$, then the
complete coupling $H$ in (\ref{eq:complete coupling}) can be written
as the \emph{reduced coupling} 
\begin{equation}
R=\left(R_{x}^{k}:k\in K,x\in\alpha^{k}\right),\label{eq:reduced coupling}
\end{equation}
such that
\begin{equation}
R_{\phi}=\left(R_{x}^{k}:k\in K,\phi\left(k\right)=x\right)\sim A_{\phi}.
\end{equation}
The existence of such a reduced coupling for a given $A$ is the central
theme of the theory of selective influences (Dzhafarov, 2003; Dzhafarov
\& Kujala, 2010, 2012a-b, 2013b, in press; Kujala \& Dzhafarov, 2008;
Schweickert, Fisher, \& Sung, 2012, Ch. 10), which includes the Bell-type
theorems as special cases. Using the language of the present paper,
if $R$ exists, one can say that each $A^{k}$ is influenced only
by the input $\alpha^{k}$ that directly influences it. In other words,
there are no influences that are not direct (``no context''). Other
examples from behavioral sciences involve recent work on combination
of concepts (Aerts, Gabora, \& Sozzo, in press; Bruza, Kitto, Ramm,
\& Sitbon, 2013; for a critical overview see Wang, Busemeyer, Atmanspacher,
\& Pothos, in press, and Dzhafarov \& Kujala, in press). In quantum
physics the existence of the reduced coupling represents classical,
pre-quantum determinism; it is the foundation of all Bell-type theorems
(Basoalto \& Percival, 2003; Dzhafarov \& Kujala, 2012a).

We know, however, that Bell-type inequalities are violated in quantum
physics. This leads us to explore alternatives to the assumption (\ref{eq:identity for reduced})
and to the ensuing existence of a reduced coupling. This can be done
by allowing $C_{x}^{k}$ in (\ref{eq:example connection}) to be different
from an identity coupling. The random variable $C_{x}^{k}$ is called
a \emph{connection}. If its distribution is posited, we constrain
the complete coupling (\ref{eq:complete coupling}) not just by (\ref{eq:condition for complete coupling}),
but also by its consistency with this connection:
\begin{equation}
H_{x}^{k}=\left(H_{\phi}^{k}:\phi\in\Phi,\phi\left(k\right)=x\right)\sim C_{x}^{k}.
\end{equation}
With this additional constraint, the coupling $H$ need not exist.

Generalizing, let $I$ be a subset of $K$ other than empty set and
$K$ itself. Then the \emph{$\left(I,\tau\right)\textnormal{-}$connection}
is defined as a random variable
\begin{equation}
C_{\tau}^{I}=\left(C_{\phi}^{I}:\phi\in\Phi,\phi|I=\tau\right)\label{eq:connection}
\end{equation}
such that for $\phi|I=\tau$,
\begin{equation}
C_{\phi}^{I}\sim A_{\phi}^{I}=\left(A_{\phi}^{k}:k\in I\right).
\end{equation}
Recall that $\phi|I=\tau$ is the restriction of the treatment on
a subset of its indices.%
\footnote{Strictly speaking, this notation makes the upper index $I$ in $C_{\tau}^{I}$
redundant. But it is convenient as it allows one to abridge the presentation
of $\tau$. Thus, if $K=\left\{ 1,2,3\right\} $, $I=\left\{ 1,3\right\} $,
$\phi\left(1\right)=x$, $\phi\left(3\right)=y$, then a strict reading
of $C_{\tau}^{I}$ is $C_{\left\{ \left(1,x\right),\left(3,y\right)\right\} }^{\left\{ 1,3\right\} }$,
but it is naturally abridged into $C_{x,y}^{1,3}$, which seems more
convenient than $C_{\left\{ \left(1,x\right),\left(3,y\right)\right\} }$.
Note that our opening example of a connection, $C_{x}^{k}$, is an
abridged form of $C_{\left\{ \left(k,x\right)\right\} }^{\left\{ k\right\} }$.%
} Note that the components of a given $C_{\tau}^{I}$ are jointly distributed,
but if \textcolor{black}{$\left(I,\tau\right)\not=\left(I',\tau'\right)$,}
$C_{\tau}^{I}$ and $C_{\tau'}^{I'}$ are stochastically unrelated.

Given a sequence of outputs $A$ in (\ref{eq:A}), denote the sequence
of the connections $C_{\tau}^{I}$ for all $I$ and $\tau$ by $C_{A}$
(not a random variable). Assume that the distributions of all these
connections are known. Then one can ask whether a complete coupling
$H$ for $A$ is consistent with all connections in $C_{A}$, that
is, whether in addition to (\ref{eq:condition for complete coupling})
$H$ also satisfies, for any $I\in2^{K}-\left\{ \emptyset,K\right\} $
and any $\tau\in\prod_{k\in I}\alpha^{k}$, 
\begin{equation}
H_{\tau}^{I}=\left(H_{\phi}^{I}:\phi\in\Phi,\phi|I=\tau\right)\sim C_{\tau}^{I},
\end{equation}
where
\begin{equation}
H_{\phi}^{I}=\left(H_{\phi}^{k}:k\in I\right).
\end{equation}
If this is true, then $H$ is called an \emph{Extended Joint Distribution
Sequence} (EJDS) for $\left(A,C_{A}\right)$. This notion is a generalization
of the Joint Distribution Sequence (or ``Joint Distribution Criterion
set'') that coincides with the reduced coupling (\ref{eq:reduced coupling})
in the theory of selective influences (Dzhafarov \& Kujala, 2010,
2012a, 2013b). It is obtained from EJDS by requiring that all connections
be identity ones, that is, for any $\phi,\phi'$ in (\ref{eq:connection}),
\begin{equation}
\Pr\left(C_{\phi}^{I}=C_{\phi'}^{I}\right)=1.
\end{equation}

\section{Characterizing Contextuality}

The notion of an EJDS can be used to characterize contextuality in
relation to constraints imposed on the outputs of a system. Suppose
that it is known that the outputs $A$ taken across all allowable
treatments in (\ref{eq:A}) satisfy a certain property $\mathcal{P}\left(A\right)$.
This property may be described by certain equations and inequalities
relating to each other parameters of the outputs, such as Bell-type
inequalities, or Cirelson-Landau's quantum inequalities (see below).
One should investigate then the set of possible $C_{A}$ in relation
to this property $\mathcal{P}\left(A\right)$.

To understand this better, let us consider a simple example of $A$.
Let $K$ be $\left\{ 1,2\right\} $, the sequence of inputs $\left(\alpha^{k}:k\in K\right)$
be $\left(\alpha^{1}=\left\{ 1,2\right\} ,\alpha^{2}=\left\{ 1,2\right\} \right)$,
the sequence of allowable treatments be $\Phi=\alpha^{1}\times\alpha^{2}$,
and the sequence of outputs be $A=\left(\left(A_{ij}^{1},A_{ij}^{2}\right):i,j\in\left\{ 1,2\right\} \right)$
(where each subscript $ij$ represents the treatment $\left(i,j\right)$).
The diagram of direct influences is assumed to be
\begin{equation}
\xymatrix{\alpha^{1}\ar[d] & \alpha^{2}\ar[d]\\
A^{1} & A^{2}
}
\end{equation}
The only choices of $I\subset K$ here other than $\emptyset$ and
$K$ are the singletons $\left\{ 1\right\} $ and $\left\{ 2\right\} $,
so the only four connections are, for $i\in\left\{ 1,2\right\} $,
\begin{equation}
C_{i}^{1}=\left(C_{i1}^{1},C_{i2}^{1}\right),C_{i}^{2}=\left(C_{1i}^{2},C_{2i}^{2}\right),\label{eq:connections in example}
\end{equation}
where $C_{ij}^{k}\sim A_{ij}^{k}$ for all $i,j,k\in\left\{ 1,2\right\} $.
Recall that the logic of forming $C_{i}^{1}=\left(C_{i1}^{1},C_{i2}^{1}\right)$
is that $A_{i1}^{1}$ and $A_{i2}^{1}$, while they are recorded at
different treatments, $\left(i,1\right)$ and $\left(i,2\right)$,
share the same pattern of direct influences, namely, both are directly
influenced by the value $i$ of $\alpha^{1}$ (in our general notation,
$\phi|\left\{ 1\right\} =\left(i\right)$). So if their joint distribution
is described by anything other than $\Pr\left(C_{i1}^{1}=C_{i2}^{1}\right)=1$,
we can speak of indirect, contextual influences. The situation with
$C_{i}^{2}$ is analogous. The complete coupling for $A$ here is
the 8-vector 
\begin{equation}
H=\left(H_{ij}^{1},H_{ij}^{2}:i,j\in\left\{ 1,2\right\} \right).
\end{equation}
Assume that each $A_{ij}^{k}$ (hence also $H_{ij}^{k}$ in the complete
coupling, $i,j,k\in\left\{ 1,2\right\} $) is a binary random variable
with equiprobable outcomes +1 and -1. Then $A$ is represented by
four probabilities $p=\left(p_{11},p_{12},p_{21},p_{22}\right)$,
where 
\begin{equation}
p_{ij}=\Pr\left[A_{ij}^{1}=+1,A_{ij}^{2}=+1\right]=\Pr\left[H_{ij}^{1}=+1,H_{ij}^{2}=+1\right].\label{eq:represent A}
\end{equation}
One prominent situation encompassed by this example is the Bohmian
version of the EPR paradigm involving two spin-$\nicefrac{1}{2}$
particles with two settings (spatial directions) per particle. As
examples of a constraint $\mathcal{P}\left(A\right)$ consider the
Bell/CH/Fine inequalities (Bell, 1964; Clauser \& Horn, 1974; Fine,
1982) 
\begin{equation}
0\leq p_{ij}+p_{ij'}+p_{i'j'}-p_{i'j}\leq1\label{eq:CH/Fine}
\end{equation}
and Cirel'son's (1980) inequalities 
\begin{equation}
\frac{1-\sqrt{2}}{2}\leq p_{ij}+p_{ij'}+p_{i'j'}-p_{i'j}\leq\frac{1+\sqrt{2}}{2},\label{eq:Cirel'son}
\end{equation}
where $i,j\in\left\{ 1,2\right\} $, $i'=3-i$, $j'=3-j$ (so each
expression contains four double-inequalities). The Bell/CH/Fine inequalities
are known to be necessary and sufficient for the existence of a classical
explanation for the EPR paradigm in question (Fine, 1982), whereas
the Cirel'son inequalities are necessary for the existence of a quantum
mechanical explanation (Landau, 1987). 

One question to pose about the connections is: what is the set of
all $C_{A}$ such that whenever $\mathcal{P}\left(A\right)$ is satisfied,
an EJDS for $\left(A,C_{A}\right)$ exists? We call any connection
belonging to this $C_{A}$ a \emph{fitting connection} for $\mathcal{P}\left(A\right)$.
A question can also be posed about the opposite implication: what
is the set of all $C_{A}$ such that whenever an EJDS for $\left(A,C_{A}\right)$
exists, $\mathcal{P}\left(A\right)$ is satisfied? We call any connection
in this $C_{A}$ a \emph{forcing connection} for $\mathcal{P}\left(A\right)$.
In our example, $C_{A}$ is the sequence of four connections $C_{i}^{k}$
in (\ref{eq:connections in example}), and they are uniquely characterized
by the 4-vector $\varepsilon=\left(\varepsilon_{1}^{1},\varepsilon_{2}^{1},\varepsilon_{1}^{2},\varepsilon_{2}^{2}\right)$,
where
\begin{equation}
\varepsilon_{i}^{1}=\Pr\left[C_{i1}^{1}=+1,C_{i2}^{1}=+1\right],\varepsilon_{i}^{2}=\Pr\left[C_{1i}^{2}=+1,C_{2i}^{2}=+1\right].
\end{equation}
Hence the complete coupling $H$, in order to be an EJDS for $\left(A,C_{A}\right)$,
should satisfy not only (\ref{eq:represent A}), but also
\begin{equation}
\Pr\left[H_{i1}^{1}=+1,H_{i2}^{1}=+1\right]=\varepsilon_{i}^{1},\Pr\left[C_{1i}^{2}=+1,C_{2i}^{2}=+1\right]=\varepsilon_{i}^{2},
\end{equation}
for $i\in\left\{ 1,2\right\} $. 

To describe the fitting and forcing connections for our example, it
is convenient to introduce the following abbreviations:
\begin{equation}
\begin{array}{c}
s_{0}=\max\left\{ \pm\left(\varepsilon_{1}^{1}-\nicefrac{1}{4}\right)\pm\left(\varepsilon_{1}^{2}-\nicefrac{1}{4}\right)\pm\left(\varepsilon_{2}^{1}-\nicefrac{1}{4}\right)\pm\left(\varepsilon_{2}^{2}-\nicefrac{1}{4}\right):\textnormal{\# of + signs is even}\right\} ,\\
s_{1}=\max\left\{ \begin{array}{l}
\pm\left(\varepsilon_{1}^{1}-\nicefrac{1}{4}\right)\pm\left(\varepsilon_{1}^{2}-\nicefrac{1}{4}\right)\pm\left(\varepsilon_{2}^{1}-\nicefrac{1}{4}\right)\pm\left(\varepsilon_{2}^{2}-\nicefrac{1}{4}\right):\textnormal{\# of + signs is odd}\end{array}\right\} .
\end{array}
\end{equation}
It turns out (Dzhafarov and Kujala, 2013a) that the sets of fitting
connections for the Bell/CH/Fine and Cirel'son inequalities are described
by, respectively,
\begin{equation}
s_{1}\le\nicefrac{1}{2},
\end{equation}
and
\begin{equation}
s_{0}\le\frac{3-\sqrt{2}}{2},s_{1}\le\nicefrac{1}{2}.
\end{equation}
This means that if $p$ satisfies (\ref{eq:CH/Fine}), then any $\varepsilon$
with $s_{1}\le\nicefrac{1}{2}$ is compatible with it, that is, this
$p$ and this $\varepsilon$ can be embedded in the same EJDS $H$.
If $p$ satisfies (\ref{eq:Cirel'son}), the set of $\varepsilon$
compatible with it is more narrow: they should additionally satisfy
$s_{0}\le\frac{3-\sqrt{2}}{2}$. Both sets include, of course, the
vector $\varepsilon=\left(0,0,0,0\right)$, which represents no-contextuality
and corresponds to the reduced coupling $R$ in (\ref{eq:reduced coupling}).

The sets of forcing connections for the Bell/CH/Fine and Cirel'son
inequalities are described by, respectively,
\begin{equation}
s_{0}=1,
\end{equation}
and
\begin{equation}
s_{0}\geq\frac{3-\sqrt{2}}{2}.
\end{equation}
The set of $\varepsilon$ such that $s_{0}=1$ consists of $\varepsilon=\left(0,0,0,0\right)$,
$\varepsilon=\left(\nicefrac{1}{2},\nicefrac{1}{2},\nicefrac{1}{2},\nicefrac{1}{2}\right)$,
and vectors with two zeros and two $\nicefrac{1}{2}$'s. All of them
represent no-contextuality, with +1 and -1 interpreted differently
in different connections. Only if $\varepsilon$ is one of these vectors,
$p$ must satisfy the Bell/CH/Fine inequalities in order to be compatible
with it. In other words, such an $\varepsilon$ and no other ``forces''
$p$ to satisfy these inequalities. The class of $\varepsilon$ that
force $p$ to satisfy the Cirel'son inequalities should include these
$\varepsilon$ because every $p$ satisfying (\ref{eq:CH/Fine}) also
satisfies (\ref{eq:Cirel'son}). But there are other $\varepsilon$,
all those with $s_{0}\geq\frac{3-\sqrt{2}}{2}$, that too are compatible
with $p$ only if they satisfy the Cirel'son inequalities.

The above serves only as a demonstration of how one could characterize
the constraints imposed on outputs (by a theory or empirical generalizations)
through the connections compatible with them, in the sense of being
embeddable in the same coupling. It should be noted, however, that
connections generally do not characterize couplings uniquely. This
opens ways for constructing qualified Kolmogorovian models more general
than the one presented in this paper.

\section{Conclusion}

We have shown that the classical, if qualified, Kolmogorovian probability
theory is not synonymous with the classical explanation of the input-output
relations (especially, in the entanglement paradigm of quantum physics).
The latter, since Bell's (1964) pioneering work, has been understood
as the existence of a single sample space for all outputs when each
output is indexed (identified) only by the inputs that directly influence
it. In the qualified Kolmogorovian approach, however, this is only
one of a potential infinity of possibilities. Different treatments
(combinations of values of all inputs) correspond to stochastically
unrelated random variables, and these can be coupled in many different
ways. Only one of these ways, with identity connections, corresponds
to John Bell's single sample space.

\subsubsection*{Acknowledgments.}

This research has been supported by the NSF grant SES-1155956. We
are grateful to Jerome Busemeyer of Indiana University for critically
reviewing this paper.

\label{references}


\begin{thebibliography}{10}
\bibitem{key-20}Aerts, D., Gabora, L., \& Sozzo, S.: Concepts and
their dynamics: A quantum-theoretic modeling of human thought. Topics
in Cognitive Science\emph{ }(in press) 

\bibitem{key-4}Basoalto, R.M., Percival, I.C.: BellTest and CHSH
experiments with more than two settings. Journal of Physics A: Mathematical
\& General, 36, 7411\textendash{}7423 (2003)

\bibitem{key-2}Bell, J.: On the Einstein-Podolsky-Rosen paradox.
Physics 1, 195-200 (1964)

\bibitem{key-1}Bruza, P.D. , Kitto, K., Ramm, B.J., Sitbon. L.: A
probabilistic framework for analysing the compositionality of conceptual
combinations. arXiv:1305.5753 (2013)

\bibitem{key-17}Cirel'son, B.S.: Quantum generalizations of Bell's
inequality. Letters in Mathematical Physics 4, 93--100 (1980)

\bibitem{key-22}Clauser, J.F., Horne, M.A.: Experimental consequences
of objective local theories\emph{. }Physical Review D\emph{ }10, 526-535
(1974)

\bibitem{Dzh2003c}Dzhafarov, E.N.: Selective influence through conditional
independence. Psychometrika 68, 7-26 (2003)

\bibitem{DzhKuj2010}Dzhafarov, E.N., Kujala, J.V.: The Joint Distribution
Criterion and the Distance Tests for selective probabilistic causality.
Frontiers in Quantitative Psychology and Measurement 1:151 doi: 10.3389\slash{}fpsyg.2010.0015
(2010)

\bibitem{DzhKuj2012}Dzhafarov, E.N., Kujala, J.V.: Selectivity in
probabilistic causality: Where psychology runs into quantum physics.
Journal of Mathematical Psychology 56, 54-63 (2012a)

\bibitem{key-23}Dzhafarov, E.N., Kujala, J.V.: Quantum entanglement
and the issue of selective influences in psychology: An Overview.
Lecture Notes in Computer Science 7620, 184-195 (2012b)

\bibitem{key-9}Dzhafarov, E.N., \& Kujala, J.V.: All-possible-couplings
approach to measuring probabilistic context. PLoS ONE 8(5): e61712.
doi:10.1371/journal.pone.0061712 (2013a)

\bibitem{key-13}Dzhafarov, E.N., Kujala, J.V.: Order-distance and
other metric-like functions on jointly distributed random variables.
Proceedings of the American Mathematical Society, 141, 3291-3301.
(2013b)

\bibitem{key-8}Dzhafarov, E.N., \& Kujala, J.V.: On selective influences,
marginal selectivity, and Bell/CHSH inequalities. Topics in Cognitive
Science (in press)

\bibitem{Fine1982b}Fine, A.: Hidden variables, joint probability,
and the Bell inequalities.\emph{ }Physical Review Letters 48, 291-295
(1982)

\bibitem{key-2}\label{enu:A.-Yu.-Khrennikov,}Khrennikov, A.Yu.:
Bell-Boole inequality: Nonlocality or probabilistic incompatibility
of random variables? Entropy 10, 19-32 (2008a)

\bibitem{key-3}\label{enu:A.-Yu.-Khrennikov,-1}Khrennikov, A.Yu.:
EPR\textendash{}Bohm experiment and Bell\textquoteright{}s inequality:
Quantum physics meets probability theory. Theoretical and Mathematical
Physics\textbf{ }157, 1448\textendash{}1460 (2008b)

\bibitem{key-4}\label{enu:A.-Yu.-Khrennikov,-2}Khrennikov, A.Yu.:
Contextual Approach to Quantum Formalism. Berlin: Springer (2009)

\bibitem{KujDzh2008}Kujala, J.V., Dzhafarov, E.N.: Testing for selectivity
in the dependence of random variables on external factors. Journal
of Mathematical Psychology 52, 128--144 (2008)

\bibitem{key-18}Landau, L. J.: On the violation of Bell's inequality
in quantum theory. Physical Letters A\textbf{ }120, 54--56 (1987)

\bibitem{key-6}\label{enu:F.-Laudisa,-Contextualism}Laudisa, F.:
Contextualism and nonlocality in the algebra of EPR observables. Philosophy
of Science 64, 478-496 (1997) 

\bibitem{key-15}Schweickert, R., Fisher, D.L., \& Sung, K.: Discovering
Cognitive Architecture by Selectively Influencing Mental Processes\emph{.}
New Jersey: World Scientific (2012)

\bibitem{key-21}Thorisson, H.\emph{: }Coupling, Stationarity, and
Regeneration\emph{.} New York: Springer (2000)

\bibitem{key-3}Wang, Z., Busemeyer, J. R., Atmanspacher, H., \& Pothos,
E. (in press). The potential of using quantum theory to build models
of cognition. Topics in Cognitive Science.\end{thebibliography}
\end{document}